# Inversed Vernier effect based single-mode laser emission in coupled microdisks

Meng Li[1], Nan Zhang[1], Kaiyang Wang[1], Jiankai Li[1], Shumin Xiao[2,†], Qinghai Song[1,3,*]

1. Integrated Nanoscience Lab, Department of Electrical and Information Engineering, Harbin Institute of Technology, Shenzhen, China, 518055
2. Department of Material Science and Engineering, Harbin Institute of Technology, Shenzhen, China, 518055

3. State Key Laboratory of Tunable Laser Technology, Harbin Institute of Technology, Harbin, China, 158001

[*] qinghai.song@hitsz.edu.cn, [†]shuminxiao@gmail.com

**Abstract:**

Recently, on-chip single-mode laser emission has attracted considerable research attention due to its wide applications. While most of single-mode lasers in coupled microdisks or microrings have been qualitatively explained by either Vernier effect or inversed Vernier effect, none of them have been experimentally confirmed. Here, we studied the mechanism for single-mode operation in coupled microdisks. We found that the mode numbers had been significantly reduced to nearly single-mode within a large pumping power range from threshold to gain saturation. The detail laser spectra showed that the largest gain and the first lasing peak were mainly generated by one disk and the laser intensity was proportional to the frequency detuning. The corresponding theoretical analysis showed that the experimental observations were dominated by internal coupling within one cavity, which was similar to the recently explored inversed Vernier effect in two coupled microrings. We believe our finding will be important for understanding the previous experimental findings and the development of on-chip single-mode laser.

## Introduction

A traditional laser system consists of gain materials and a laser resonator [1]. Despite of the ultrasmall cavities [2], most of resonators support a lot of longitude modes with small free spectral range (FSR) due to their lengths of round trips [3-5]. Consequently, multimode lasers are usually generated without utilizing external coupling or adding intra-cavity dispersive elements, especially in the on-chip microlasers. To improve the monochromacity, several techniques have been developed in past few years [6-17]. The application of Vernier effect in coupled cavities is a prominent example. Vernier effect is a well-known technique in passive systems to extend the FSR of bandpass filters [18]. Due to the destructive interference, the lasing modes, except for the coupled resonances, are significantly suppressed. Based on this idea, several groups have successfully demonstrated single-mode laser emission in dye-doped micro-lasers or quantum cascade lasers [10-17]. However, only the interferometer based single-mode laser has been verified to be generated by Vernier effect [10-12]. The others such as coupled microdisks or microrings are not well explained and their corresponding mechanisms are often argued.

Very recently, a different mechanism, which is named as inversed Vernier effect, has been proposed to explain the suppression of mode numbers in size-mismatched coupled cavities [19]. Compared with the Vernier effect, the mode spacing under inversed Vernier effect can also be extended by the different FSRs in two cavities following the same equation [10, 18, 19]. Meanwhile, the amplification or threshold is also dependent on the frequency detuning $\Delta\omega$ or coupling constant. The only difference is that the thresholds of modes with largest frequency detuning are close to the maximum values. In this sense, the inversed Vernier effect can suppress the number of modes as significantly as the Vernier effect and gives the same FSR. Consequently, the qualitative analysis based on the extended FSR cannot distinguish these two effects and thus the exact mechanism for single-mode laser emission in coupled microdisks or microrings is still in the debating. In this article, we experimentally studied the suppression of mode numbers in size-mismatched coupled microdisks. By comparing the laser spectra in coupled cavities, we found that the single-mode operation was mainly generated by mode coupling in one microdisk. This new mechanism is similar to the inversed Vernier effect. But it is formed within one circular microdisk instead of the coupled microdisks.

## Results and Discussions

### Theoretical analysis and numerical simulation

To well understand the suppression of lasing modes, we have theoretically studied the possible mode interaction by a toy model. In general, both the Vernier effect and inversed Vernier effect are based on the interaction between two resonances thus can be quantum mechanically described by a 2×2 matrix [20-24]

$$\mathrm{H} = \begin{bmatrix} E_1 & J \\ J & E_2 \end{bmatrix} \tag{1}$$

Here $E_{1,2}$ are the energy of states away from mode coupling and $J^2$ is the coupling constant. Considering the main interaction via the field distributions, the coupling factor J can be simply considered as a real number [24]. Then the eigenvalues of Eq. (1) are expressed as

$$\mathrm{E}_{\pm} = \frac{\mathrm{E}_1+\mathrm{E}_2}{2} \pm \sqrt{\left(\frac{\mathrm{E}_1-\mathrm{E}_2}{2}\right)^2 + J^2} \tag{2}$$

Figures 1(a) and 1(b) illustrate the real and imaginary parts of energy as a function of frequency detuning (Δ) without considering the nonlinearity in laser systems. The amplifications of lasing modes were simply introduced from the imaginary parts of energy. We can see that the real parts of energy experienced a repulsion around Δ = 0 and their imaginary parts crossed. Most importantly, max(Im ($E_1$), Im($E_2$)) reduced round the crossing point. All these results are very typical phenomena of avoided resonance crossing. For a lasing system, the reduction of max(Im ($E_1$), Im($E_2$)) means that the amplification of lasing mode is reduced. In this sense, it is easy to know that the modes around mode coupling should have higher thresholds and hard to be excited in lasing experiment. In this sense, the mode coupling has the possibility to generate single mode laser emission, which is similar to the inversed Vernier effect [19]. Moreover, for the case of internal coupling with a real J, similar phenomena have also been observed when the weak coupling happened. Thus we know that both strong coupling and weak coupling with real coupling constant J can induce the inversed Vernier effects. It is also worth to note that the model in Eq. (1) is valid for the interaction between two states. It is not necessary to be restricted in coupled systems. In this sense, the inversed Vernier effect can be extended beyond the scope of coupled microdisks or microrings.

Based on the theoretical analysis, we then tested above model numerically. In optical microcavities, the resonances and their frequencies play the roles of states and their energies. In the numerical calculation, the gain was also introduced from the imaginary parts of refractive index by setting $n_L = n_R^* = n_0 + n''i$. Here $n_0$ was set as $n_0 = 1.56$ and $n'' = - 0.001$ to mimic the experimental conditions of right-pumping. The radius of left cavity was fixed at 5 μm, whereas the size of right cavity varided between 5.09 – 5.14 μm. The calculated results have been summarized and shown in Fig. 2. Similar to the theoretical prediction from Eq. (2), we can see the frequencies of resonances approached first and then repelled one another (see Fig. 2(a)). The corresponding β of two modes crossed around $R_R$ = 5.115 μm and the gains of modes away from the crossing point are much higher (see Fig. 2(b)).

The corresponding field patterns of modes marked as 1-6 in Fig. 2(a) are shown in Figs. 2(c)-2(h). We can see that the field patterns of two resonances exchanged after the crossing point and thus confirmed the strong coupling between them. When the resonances are far away from crossing point at $R_R$ = 5.095 μm, mode-1 was mainly confined within the gain cavity (Fig. 2(c)) and mode-2 was localized within two cavities (Fig. 2(d)). As the other cavity was defined as absorption in the numerical model, mode-2 had much larger loss than mode-1. This can also been seen from their corresponding β values in Fig. 2(b). When two resonances approached the crossing point, the corresponding field patterns showed that they were strongly mixed (see Figs. 2(e) and 2(f)). Thus the hybrid resonances shall experience two kinds of orbits of mode-1 and mode-2. Then the larger amplification was reduced and the smaller amplification was increased around the ARC due to the increased field distribution in the lossy cavity. And the modes around ARC had much larger thresholds and were not easily to be excited. All these results are consistent with the theoretical prediction well.

## Experimental results

We then tested the theoretical analysis experimentally. The microdisk was directly fabricated in Rhodamine B doped SU8 photoresist by standard photolithography. The photography of fabricated microdisk is shown in Fig. 3(a), which consisted of two tangent cavities with radiuses $R_{1,2}$ ~ 40 μm. Due to the resolution of photolithography, the gap between two circular disks was filled with a bridge. The width of bridge is around 8.98 micron. The thickness of the microdisk was 2 μm, which was far smaller than the in-plane cavity size. The refractive index of SU8 is around 1.56. Then the sample was optically pumped by a frequency doubled Nd: YAG pulsed laser (532nm, pulse duration ~ 7ns). The laser was focused onto the sample surface by a 5x objective lens and the diameter of laser spot was 180 μm, which was large enough to cover the whole sample.

Figure 3(b) illustrates the laser spectrum from the microdisk. Here the whole disk was evenly pumped and the power density was ~22.22 μJ/mm$^2$. We can see a dominated laser peak at 637.45nm. The other lasing modes are dramatically suppressed. This is quite different from the conventional whispering gallery modes in either circular microdisk or coupled cavities. The green open squares in Fig. 3(c) clearly demonstrated the threshold curve in logscale. When the pump power was small, a broad spontaneous emission peak could be observed in the laser spectrum. And the slope of L-I curve was around 1. Once the pump power is above 15.58 $\mu J/mm^2$, the single mode laser appeared and the slope was increased to 3.43. The laser emission was saturated and the slope transited back to ~ 1 again when the pump power was further increased. All these behaviors confirmed the lasing action in microdisk very well. The corresponding laser spectra have been shown in Fig. 3(d). In additional to the transition from spontaneous emission to laser, it is more interesting to see that the single-mode laser can be observed in a wide range of pumping power between the threshold and gain saturation. The spectra at extremely high pump power showed three peaks because the intensity of main peak was saturated in the detection.

The above observations are very similar to the recent experimental reports in coupled cavities. Thus it is interesting to explore the intrinsic mechanism for the single-mode operation. To achieve enough information, we have pumped the left and right cavity individually by moving the relative position between sample and laser spot. Below, these two pumping configurations will be named as left-pumping and right-pumping. Figure 4 shows the logscale laser spectra of the sample under different pumping conditions. Here the pump power was increased to 43.47 $\mu J/mm^2$ to ensure the lasing behaviors of the sample under three pumping configurations. We can see that the coupled cavities produced multiple laser peaks when it was under left pumping (see Fig. 4(c)). Once the pumping configuration was switched to right pumping, the mode number was significantly reduced and only three peaks around 637nm could be observed (see Fig. 4(b)). More interestingly, while the laser spectrum in Fig. 4(a) contained all the laser peaks in Figs. 4(b) and 4(c), the three peaks that dominated the laser spectrum were almost the same as the ones in Fig. 4(b). This shows that the suppression of lasing modes in coupled cavities is mainly induced by one cavity instead of the coupling between two cavities. This can also be confirmed by counting the mode spacing between the modes in two cavities. Because the coupled cavities were very close, the wavelength differences between modes in two cavities were mostly around 0.05 nm. Thus the suppression of mode number in coupled cavity is different from the previous studies about Vernier effect and inversed Vernier effect.

To explore the mechanism of mode suppression in right pumping, we have increased the pump power to excite more lasing modes and achieve more spectral information. One example is depicted in Fig. 5(a). As the other modes were still much smaller, here the laser spectrum was plotted in logscale. We can see that the three main peaks belong to one set of modes, which are named as seires-1 and marked by red arrows in Fig. 5(a). Meanwhile, we can also see another set of modes nearby, which are named as series-2 and marked with blue arrows in Fig. 5(a). From their individual mode spacing, these two sets of modes correspond to the whispering gallery modes within circular cavity. We note the lasing modes confined in two cavities are not considered here because the left cavity was not pumped and functioned as a pure absorber in this experiment. The interesting phenomena happened at the relative wavelength shift between two sets of lasing modes. As shown in Fig. 5(b), we can see that the mode spacing between two sets of modes increased first and then decreased. If we consider the possible interaction between two sets of resonances, Fig. 5(b) demonstrated that the mode coupling was smallest at 637.4nm, where the maxima laser peak in Figs. 3-5 appeared.

According to the dependence of maxima laser on the wavelength shift, the suppression of lasing modes in a single cavity is very similar to above theoretical and numerical analysis. Then the suppression of mode number in coupled cavities can be explained. As the laser thresholds of modes in seires-2 are much larger than those of modes in series-1, the resonances in series-2 are more lossy than series-1. When the modes in series-1 couple with the modes in series-2, the waves confined within the orbits of series-1will partially transfer to the orbits of series-2 and thus experience more loss. In this sense, the mode with least coupling to modes in series-2 has the lowest loss and smallest threshold. Thus only the resonances far away from the coupling point can be excited and the mode numbers in laser systems has been dramatically decreased. All these observation are consistent with the predictions in Fig. 1 and 2 well.

Meanwhile, as the intensities of modes in the other cavity didn't fluctuate as significantly as right-pumping, the strongest lasing modes in right-pumping should also be the strongest under evenly pumping. Thus the total lasing mode number had been suppressed under both right-pumping and evenly-pumping. Moreover, the laser peaks far away from the maxima wavelength detuning should be close to the avoided resonance crossings. According to the predictions in Fig. 1(b), the amplifications of two coupled resonances should be similar. This information could also be observed in Fig. 5(a). In the wavelength range below 635nm, we can see that two sets of resonances had similar intensities. This can be a further proof for the mode coupling in the laser system.

## Conclusion

In summary, we have experimentally studied the lasing actions within coupled microdisks. By comparing the results under three types of pumping conditions, we found the single-mode operation in coupled cavities were mainly induced by the mode coupling within one cavity. The corresponding theoretical analysis and numerical calculation also proved our qualitative model well. Our results demonstrated that the inversed Vernier effect could induce single-mode operation for the first time and successfully extend the typical inversed Vernier effect from coupled microdisks or micro-rings to a single cavity as well. We believe that our finding can be nice way to generate single mode laser emission and might also be used to explain the previous experimental observations well.

## Method

### Fabrication and optical characterization

The microdisk in the present study was made by commercially available GM 1050 doped with 0.85%(by weight) RhB. We choose commercial glass sheet as the substrate. The photoresist was spin-coated on the substrate with 4000r per minute and had thickness 2 micron. After 8 minutes ultraviolet exposure under the density 1000mW/cm$^2$, the sample was developed in PGMEA for 17 seconds. The obtained microdisk consisted of two tangent circular microdisks with radiuses around 40 μm. The designed gap was partially filled with a bridge due to the resolution of photolithography. The width of bridge was 8.98 μm.

The coupled cavities was mounted on a three-dimensional translation stage and pumped by a 532nm pulsed light coming from a mode-locked Nd:YAG laser. The laser spot was focused by a 5x objective lens to a spot with diameter around 90 μm. During the experiment, the laser spot was fixed and the sample was moved around the laser spot to form three pumping configurations. The emitted lasers were collected by a lens and coupled trough a multimode fiber to a spectrometer and a CCD camera (Princeton instrument).  The far field patterns were directly recorded with a camera by placing a screen at the far field.

## Numerical simulations

Because the thicknesses of microdisks were much smaller than their in-plane dimensions, microdisks were usually treated as two-dimensional objects by applying effective refractive indices *n*. The gain and loss of the microdisk are introduced by applying $n_L = n_0 + in''$ and $n_R = n_0 - in''$. Then the wave equations for transverse electric (TE, E is in plane) polarized modes $H_z(x, y, t) = \psi(x, y)e^{-i\omega t}$ can be replaced by the scalar wave equation

$$-\nabla^2\psi = n^2(x, y)\frac{\omega^2}{c^2}\psi \qquad (4)$$

with angular frequency ω and speed of light in vacuum *c*.

We numerically computed the TE polarized resonances by solving above equation use FEM with the RF module in COMSOL Multiphysics 3.5a. The outgoing waves are absorbed by perfect matched layer at far field, leading to quasibound states with complex eigenfrequencies ($\omega$). The wave number is defined as $\beta = \omega/c$.

## References

1. Siegman, A. E. *Lasers*. University Science Books, (1986).
2. Gu, Z. Y. et al., Photon hopping and nanowire based hybrid plasmonic waveguide and ring-resonator. *Sci. Rep.* **5**, 9171 (2015).
3. Kneissl, M. et al., Current-injection spiral-shaped microcavity disk laser diodes with unidirectional emission. *Appl. Phys. Lett.* **84**, 2485-2487 (2004).
4. Schermer, M., Bittner, S., Singh, G., Ulysse, C., Lebental, M., & Wiersig, J., Unidirectional light emission from low-index polymer microlasers. *Appl. Phys. Lett.* **106**, .101107 (2015).
5. Song, Q. H. et al., Directional laser emission from a wavelength-scale chaotic microcavity. *Phys. Rev. Lett.* **105**, 103902 (2010).
6. Feng, L., Wong, Z. J., Ma, R.-M., Wang, Y., & Zhang, X., Single-mode Laser by Parity-time Symmetry Breaking. *Science* **346**, 972-975 (2014).
7. Hodaei, H., Miri, M., Heinrich, M., Christodoulides, D. N., & Khajavikan, M., Parity-time-symmetric microring lasers. *Science* **346**, 975-978 (2014).
8. Elias, L. R., Ramian, G., Hu, J., & Amir, A., Observation of Single-Mode Operation in a Free-Electron Laser, *Phys. Rev. Lett.* **57**, 424 (1986).
9. Song, Q.H. et al., Liquid-Crystals-based Tunable High Q Directional Random Laser from a Planar Random Microcavity. *Opt. Lett.* **32**, 373 (2007).

10. Liu, P. Q., Wang, X., & Gmachl, C. F., Single-mode quantum cascade lasers employing symmetric Mach-Zehnder interferometer type cavities. *Appl. Phys. Lett.* **101**, 161115 (2012).

11. Zheng, M. C et al., Wide single-mode tuning in quantum cascade lasers with asymmetric Mach-Zehnder interferometer type cavities with separately biased arms. *Appl. Phys. Lett.* **103**, 211112 (2013).

12. Yu, Y. L., Liaw, S. K., Hsu, W. C., Shih, M. H., & Chen, N. K., Single longitudinal mode Ytterbium doped fiber lasers with large proposed tuning range. *Opt. & Quantum Electro.* **47**, s11082-014-9891-5 (2015).

13. Liu, P. Q., Wang, X., Fan, J. Y., & Gmachl, C. F., Single-mode quantum cascade lasers based on a folded Fabry-Perot cavity. *Appl. Phys. Lett.* **98**, 061110 (2011).

14. Liu, P. Q., Sladek, K., Wang, X., Fan, J. F., & Gmachl, C. F., Single-mode quantum cascade lasers employing a candy-cane shaped monolithic coupled cavity. *Appl. Phys. Lett.* **99**, 241112 (2011).

15. Shang, L, Liu, L. Y., & Xu, L., Single-frequency coupled asymmetric microcavity laser, *Opt. Lett.* **33,** 1150-1152 (2008).

16. Tu, X., Wu, X., Li, M., Liu, L. Y., and Xu, L. Ultraviolet single-frequency coupled optofluidic ring resonator dye laser. *Opt. Express* **20**, 19996-20001 (2012).

17. Song, Q. H., Cao, H., Ho, S. T., & Solomon, G. S., Near-IR subwavelength microdisk lasers. *Appl. Phys. Lett.* **94**, 061109 (2009).

18. Griffel, G., Vernier effect in asymmetrical ring resonator arrays. *Photon. Tech. Lett.* **12**, 1642-1644 (2000).

19. Ge, L., & H. E. Türeci, Inversed Vernier effect in coupled lasers. arXiv:1503.04382 (2015).

20. Heiss, W. D. Repulsion of resonance states and exceptional points. *Phys. Rev. E* **61**, 929-932 (2000).

21. Wiersig, J., Formation of long-lived, scarlike modes near avoided resonance crossings in optical microcavities. *Phys. Rev. Lett.* **97**, 253901 (2006).

22. Song, Q. H., & Cao, H., Improving optical confinement in nanostructures via external mode coupling. *Phys. Rev. Lett.* **105**, 053902 (2010).

23. Song, Q. H., Gu, Z. Y., Liu, S., & Xiao, S. M. Coherent destruction of tunneling in chaotic microcavities via three-state anti-crossings. *Sci. Rep.* **4**, 4858 (2014).

24. Magunov, A. I., Rotter, I., & Strakhov, S. I., Avoided level crossing and population trapping in atoms. *Physica E* **9**, 474-477 (2001).

## Acknowledgements

This work is supported by NSFC11204055, NSFC61222507, NSFC11374078, NCET-11-0809, Shenzhen Peacock plan under the Nos. KQCX2012080709143322 and KQCX20130627094615410, and Shenzhen Fundamental research projects under the Nos. JCYJ20130329155148184, JCYJ20140417172417110, JCYJ20140417172417096.

## Author contributions

Q.S. and S.X. designed the research. M.L., N.Z., K.W., J.L. measured the results. Q.S., S.X., and M.L. analyzed the data. Q.S. wrote the manuscript. All authors reviewed and agreed with the contents.

## Additional Information

**Competing financial interests:** The authors declare no competing financial interests.

## Figure Captions:

Fig.1: **Theoretical model inversed Vernier effect.** (**a**) and (**b**) are the real and imaginary parts of energy as a function of frequency detuning Δ. Here $E_1 = \Delta + 0.05i$, $E_2 = 0.01i$, the coupling constant J = 0.01.

Fig.2: **Numerical results about weak coupling.** (a) and (b) show the dependences of normalized frequency and the imaginary parts of propagation constant (β) on the size of right cavity. (c) and (h) are the field patterns of modes marked as 1-6 in (a). The size of left cavity is $R_R = 5$ μm, and the size of right cavity $R_L$ changes in the simulation. The refractive indices are $n_L = n_R^* = n_0 + n''I$, where $n_0$ =1.56 and n" = 0.001.

Fig.3: **Single mode lasing action in coupled cavities.** (**a**) The microscope image of the coupled cavities. (**b**) The laser spectrum of coupled microdisk under evenly pumping with pump power 22.22 μJ/mm$^2$. (**c**)_The threshold behaviors of the coupled cavities under evenly pumping (open squares), left pumping (open circles), and right pumping (open triangles). (**d**). The corresponding laser spectra under evenly pumping in (c).

Fig.4: **The laser spectra of the coupled cavities under three types of pumping configuration.** (a) evenly pumping, (b) left pumping, (c) right pumping. Here the pump power is 43.47 μJ/mm$^2$.

Fig.5: **The proof of inversed Vernier effect.** (**a**) The logscale laser spectra of right pumping. The pump power is 55.5 μJ/mm$^2$. (**b**) The frequency detuning (green squares) between two sets of modes in (a).

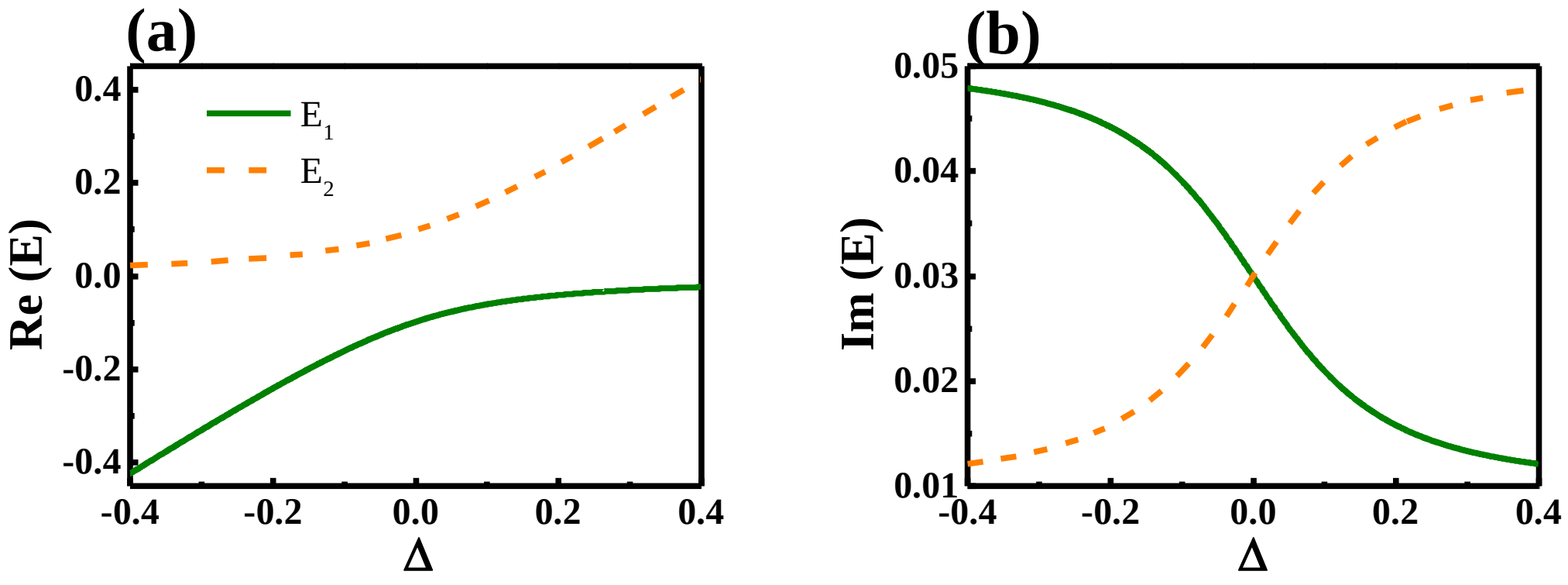

**Fig.1**

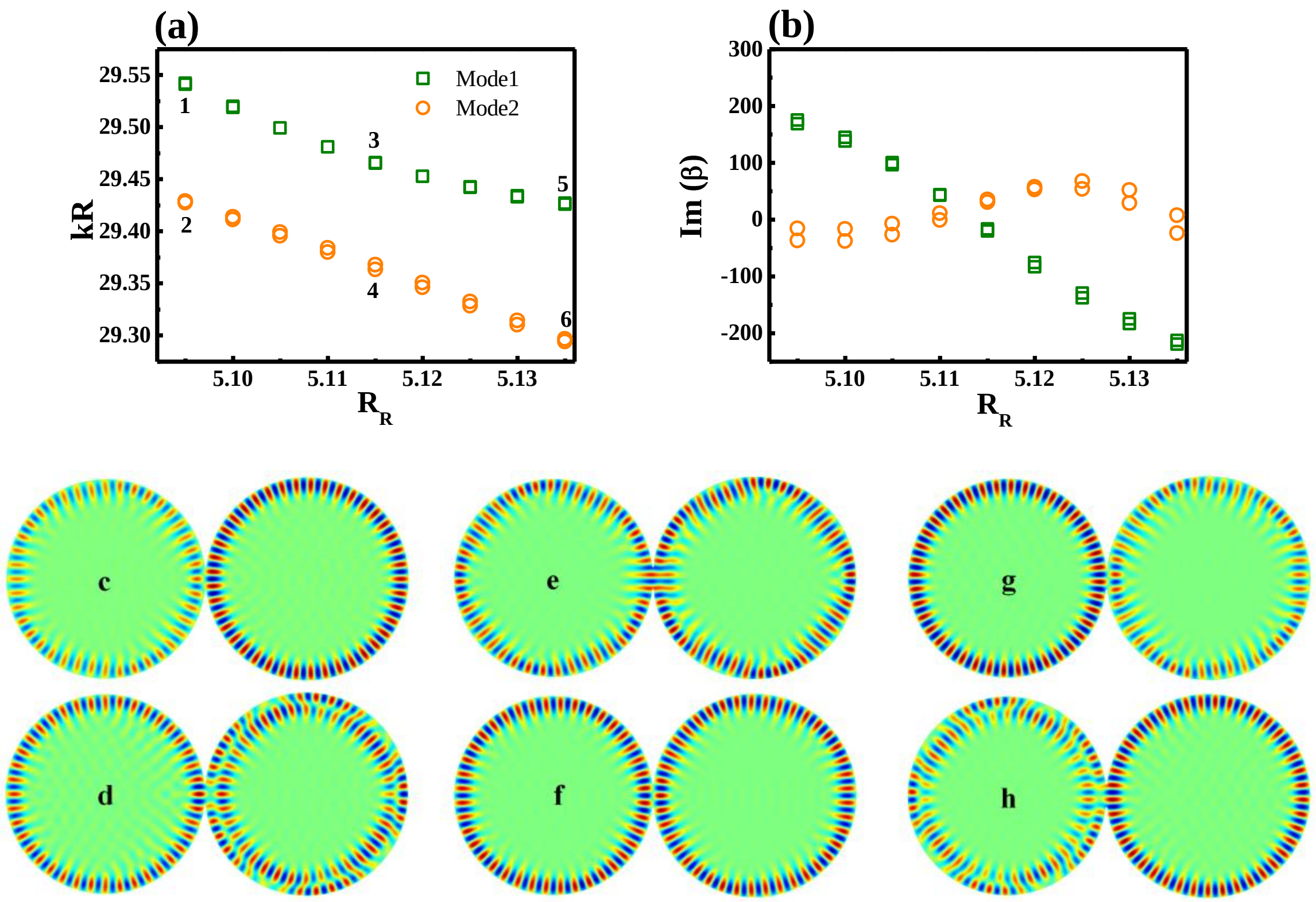

**Fig.2**

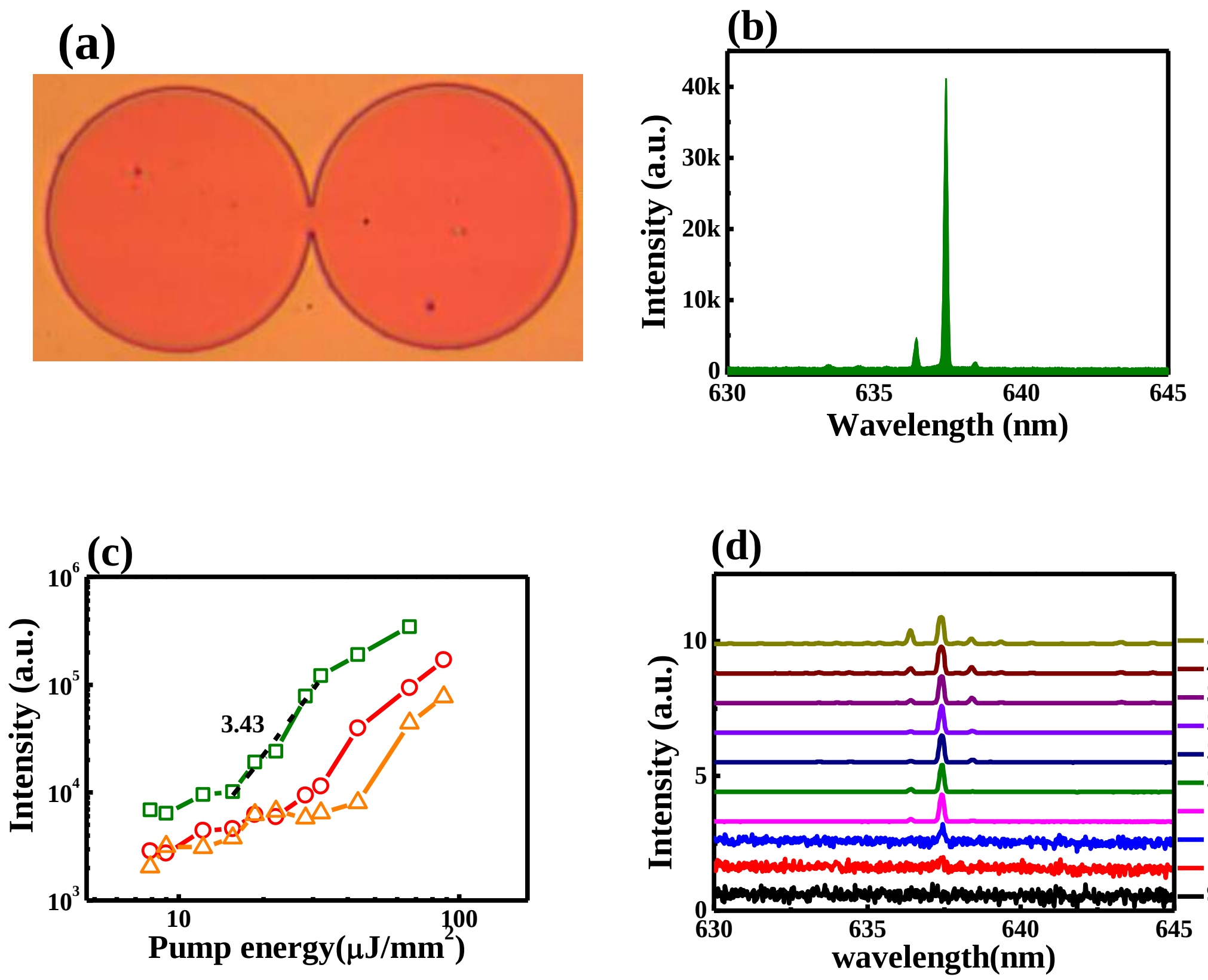

**Fig.3**

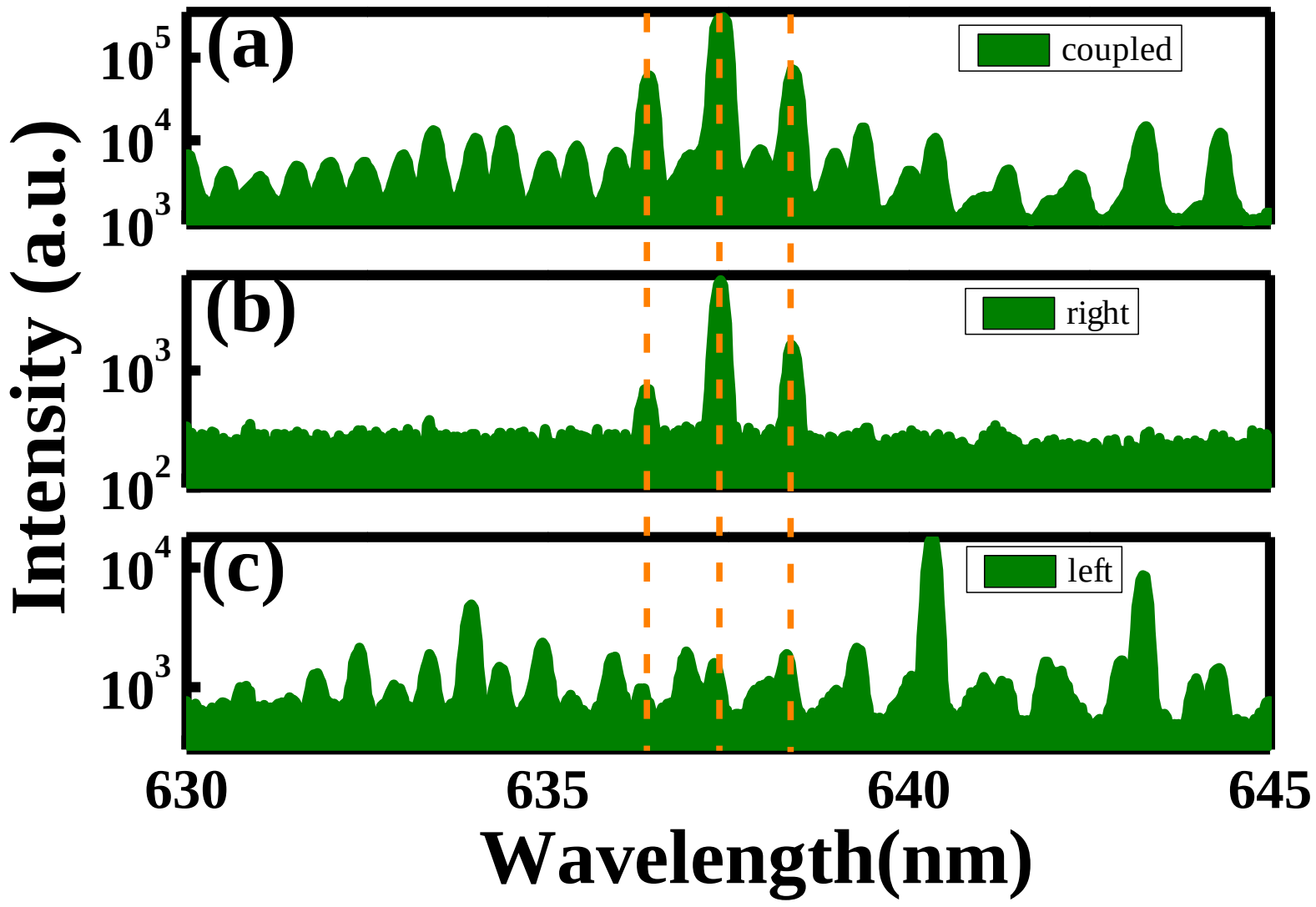

**Fig.4**

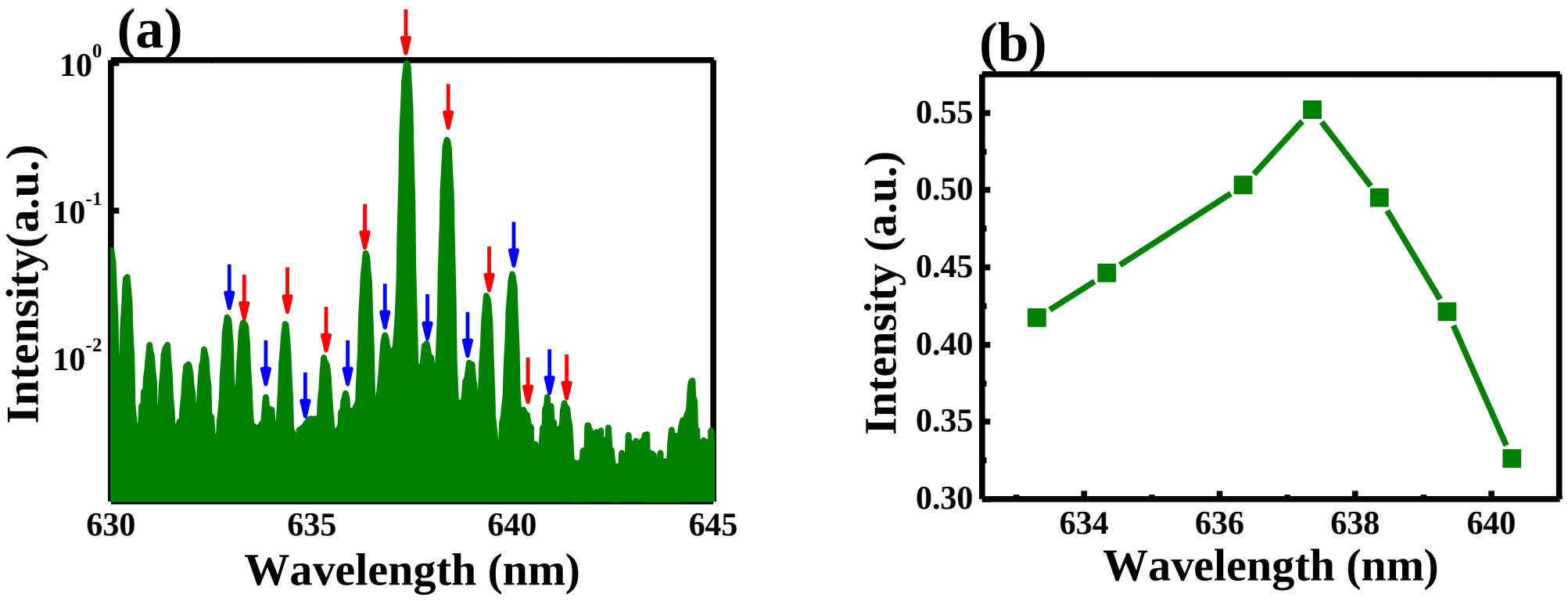

**Fig. 5**